\documentclass[traditabstract]{aa}
\usepackage{multirow}
\usepackage{graphics}
\usepackage{txfonts}
\usepackage{amssymb}
\usepackage{lscape}
\usepackage{epsfig}
\usepackage{natbib}

\begin{document}

   \title{The XXL Survey\thanks{Based on observations obtained with
        XMM-Newton, an ESA science mission with instruments and contributions
        directly funded by ESA Member States and NASA}}
 \subtitle{XXXII.  Spatial clustering of the XXL-S AGN}

    \author{M. Plionis\inst{1,2},
            L.Koutoulidis\inst{1},
            E. Koulouridis\inst{3,4},
            L. Moscardini\inst{5,6,7},
            C. Lidman\inst{8},
            M. Pierre\inst{3,4},
            C. Adami\inst{9},
            L. Chiappetti\inst{10},
            L. Faccioli\inst{3,4},
            S. Fotopoulou\inst{11},
            F. Pacaud\inst{12},
            S. Paltani\inst{13} }

 \institute{National Observatory of Athens, Lofos Nymfon, Thession, Athens 11810, Greece
    \and Physics Department, Aristotle University of Thessaloniki, Thessaloniki 54124,Greece 
    \and IRFU, CEA, Universit\'e Paris-Saclay, F-91191 Gif-sur-Yvette, France
    \and Universit\'e Paris Diderot, AIM, Sorbonne Paris Cit\'e, CEA, CNRS, F-91191 Gif-sur-Yvette, France
    \and Dipartimento di Fisica e Astronomia, Alma Mater Studiorum Universit\'a di Bologna, Via Gobetti 93/2, I-40129 Bologna, Italy
    \and INAF - Osservatorio di Astrofisica e Scienza dello Spazio, via Gobetti 93/3, I-40129 Bologna, Italy
    \and INFN, Sezione di Bologna, viale Berti Pichat 6/2, I-40127 Bologna, Italy
    \and Australian Astronomical Observatory, PO Box 915,North Ryde NSW 1670, Australia 
    \and LAM, OAMP, Universit\'e Aix-Marseille, CNRS, P\^ole de l'\'Etoile, Site de Ch\~{a}teau Gombert, 38 rue Fr\'ed\'eric
    Joliot-Curie, 13388, Marseille 13 Cedex, France
    \and INAF - IASF - Milano, Via Bassini 15, I-20133 Milano, Italy
    \and Centre for Extragalactic Astronomy, Department of Physics, Durham University, South Road, Durham, DH1 3LE, UK
    \and Argelander Institut fuer Astronomie, Universitaet Bonn, Auf dem Huegel 71, 53121 Bonn, Germany 
    \and Department of Astronomy, University of Geneva, ch. d'Ecogia 16, 1290 Versoix, Switzerland
}

\date{Received/accepted }
\abstract{The XMM-XXL Survey spans two fields of $\rm 25$ deg$^2$ each
  observed for more than 6Ms with XMM, which provided a
      sample of tens of thousands of point sources with
a flux limit of $\sim 2.2 \times 10^{-15}$ and $\sim 1.4 \times
10^{-14}$ erg s$^{-1}$ cm$^{2}$, corresponding to 50\% of the
area curve, in the soft band and hard band, respectively.
 In this paper we present
      the spatial clustering properties of $\sim 3100$ and $\sim 1900$ 
X-ray active galactic nuclei (AGNs) in the 0.5--2 and 2--10 keV bands, respectively,
      which have been 
     spectroscopically observed with the AAOmega facility. 
     This sample is 90\% 
      redshift complete down to an optical magnitude limit of
     $r\lesssim 21.8$. The sources span the redshift interval
     $0<z<5.2$, although in the current analysis we limit our samples
      to $z\le 3$, with corresponding sample median values of
      $\bar{z}\simeq 0.96$ and
      0.79 for the soft band and hard band, respectively. 
     We employ the projected two-point correlation function to infer the
 spatial clustering and find a correlation length 
$r_0=7.0 (\pm 0.34)$ and 6.42$(\pm 0.42)$ $h^{-1}$ Mpc,
  respectively, for the soft- and hard-band detected sources with a slope
  for both cases of $\gamma=1.44 (\pm 0.1)$. The power-law clustering
  was detected within comoving separations of 1 and $\sim$25 $h^{-1}$ Mpc.
These results, as well as those derived in two separate redshift ranges,
provide bias factors of the corresponding AGN host dark matter halos
that are consistent with a halo mass of $\log_{10} [M_h/(h^{-1}
M_{\odot})]=13.04\pm 0.06$, confirming the results of most recent studies
based on smaller X-ray AGN samples.}


\authorrunning{M. Plionis et al.}
\titlerunning{The XXL survey - XXXII}

\maketitle

\section{Introduction}
The importance of studying the clustering pattern of active galactic nuclei (AGNs) and their
evolution stems from
the fact that it can place important constraints on the AGN triggering
mechanisms, which are still largely unknown
\citep{Alexander12}. Furthermore, it can provide 
information regarding the properties of the dark matter halos that
host AGNs. 
 This is of great importance given the strong evidence supporting an
 interactive co-evolution of black holes and their host galaxies
  \citep[e.g.][]{Magorrian98,Gultekin09,Zubovas12}.
Semi-analytical models, 
including the effects of major galaxy mergers, have been employed
 to explain the triggering mechanism for
the most luminous AGNs \citep{DiMatteo05,Hopkins06,Marulli09}. However, secular
evolution (disc instabilities or minor interactions) could also 
be at work in the 
lowest luminosity AGN regime \citep{HopkinsHernquist06,Bournaud11}.
Merger models appear to reproduce both the clustering of quasi-stellar objects
(QSOs) and the mass of dark matter halos in which they reside. 
 Observational studies lead to somewhat conflicting clustering results
 depending on the AGN selection
 \citep[e.g.][]{Coil07,Coil09,Mendez16,Maglioc17,Hale18}.  

A large number of X-ray AGN spectroscopic surveys of varying sizes
have been used to measure
the X-ray AGN spatial correlation function, providing indications of a much
stronger clustering pattern than that of optical AGNs
\citep[][]{Mulli04,Gilli05,Yang06,Gilli09,Hickox09,Coil09,Krumpe10, 
Cappelluti10,Miyaji11,Starikova11,Allevato11,Koutoulidis13,Mountrichas16,
Allevato16,Krumpe18}. The stronger clustering of X-ray AGNs 
implies that they are hosted by more massive dark matter (DM) halos than 
optical  QSOs \citep[e.g.][]{Koutoulidis13}, while 
its dependence on X-ray
luminosity suggests that the main accretion mode of the former is the
so-called hot halo mode \citep[e.g.][]{Fanidakis12,Fanidakis13}. 
Observational indications of varying strength have been
reported in the literature for the dependence of
AGN clustering on luminosity \citep{Plionis08,Krumpe10,
  Cappelluti10,Koutoulidis13,Fanidakis13}.

This paper is part of the long list of studies based on the 
 Ultimate XMM-Newton Extragalactic X-ray survey, or XXL
\citep[][aka XXL paper I]{Pierre16}, which is an extension of the XMM-LSS survey at the same
depth,
but covering 50\,deg$^2$ in two 25\,deg$^2$ fields in the northern and
southern hemispheres (XXL-N and XXL-S, respectively). 
While the XXL
survey was primarily designed to build a consistent sample of
galaxy clusters for cosmology \citep[][XXL paper II]{Pacaud16}, an immediate
by-product of the survey is the identification of numerous point
sources, the large majority of which are AGNs.
In particular, AGN science traditionally confronted
    with low number statistics when performing statistical studies of
    the population, receives a great benefit from the addition of about
    25 000 sources.

Here we present an analysis of the clustering properties of 
a subsample of these sources which have been spectroscopically
observed with the AAOmega multifibre spectrograph. In total $\sim 3740$
XXL-S sources have currently spectroscopic data. We note that when it is 
necessary to use an {a priori} Cosmology (e.g. to estimate distances from
redshifts), we use a flat $\Lambda$CDM model with $\Omega_m=0.3$,
$\sigma_8=0.81$, and $H_0=100 \; h$ km s$^{-1}$ Mpc$^{-1}$.

In Sect. 2 we present the basic information about the XXL survey,
the multiwavelength counterparts of the X-ray point sources, and the
follow-up spectroscopic
campaign. In section 3 we present the basic methodology used to derive
the projected correlation function and the inferred spatial clustering,
while section 4 lists the basic results and a relevant
discussion regarding the inferred X-ray AGN bias evolution and the host DM
halo mass. Finally, in section 5 we present the list of the main
conclusions of our work.

\section{The XXL point source catalogue}

\subsection{X-ray source detection}
X-ray source extraction is performed in three stages, as detailed by
\citet{pacaud06}. First, images from the three EPIC detectors are
combined and a smoothed image is obtained using a multiresolution
wavelet algorithm tuned to the  low-count Poisson regime
\citep{starck98}. Source detection is then performed on this smoothed
image via Sextractor \citep{bertin96} and a list of candidate sources
is produced. Finally, a maximum likelihood (ML) fit based on the
C$-$statistic \citep{cash79} is performed for each candidate source;
only sources with a detection likelihood from the ML fit $>$ 15 are
considered significant \citep{pacaud06}. This process is performed
separately for the soft (0.5--2 keV) and hard (2--10 keV) bands.
The subsequent band merging of the detections is detailed in 
\citet[][hereafter XXL paper XXVII]{Chiappetti18}. The point source 
sample has a flux limit of $\sim 2.2 \times 10^{-15}$ and $\sim 1.4 \times
10^{-14}$ erg s$^{-1}$ cm$^{2}$ (corresponding to 50\% of the area
curve) for the soft band and hard band, respectively (see figure 3 in
XXL paper XXVII).

\subsection{Multiwavelength counterpart assignment}
Several telescopes have targeted both the XXL-North and XXL-South
fields, either as part of an all-sky survey or due to dedicated
proposals. XXL-North holds a privileged position with respect to the
southern field since it is overlapping with one of the CFHTLS fields,
namely W1. Nevertheless, significant progress is being made with
observations also accumulating  for the southern field, which already
benefits from multiwavelength coverage, 
i.e. surveys based on the Galaxy Evolution Explorer (GALEX), 
the Visible and Infrared Survey Telescope for Astronomy (VISTA), the
Infrared Array Camera on Spitzer (IRAC), Wide-field Infrared Survey
Explorer (WISE), the Blanco Cosmology Survey (BCS), and the Dark Energy
Camera (DECam). We have retrieved all the images available for these
surveys along with the corresponding weight maps when available. All
the optical and near-infrared images have been rescaled to zero-point
30 for consistency and ease during the photometry extraction. GALEX,
IRAC, and WISE data  already have homogeneous zero-points (per
survey), and there is thus no need for rescaling.

\subsection{Counterpart association}
To associate X-ray sources with potential counterparts
in other wavelengths,  we first obtained multiwavelength counterpart sets, by
positionally matching the individual primary photometric catalogues
(one per survey per band: 28 in XXL-N and 19 in XXL-S) using a radius
among them of 0.7${''}$ (or 2${''}$ for GALEX, IRAC, and WISE) as described in
\citet[][hereafter XXL paper VI]{Fotopoulou16}. 

 We then computed the likelihood ratio estimator \citep{Sutherland92},
 which has also been used in other studies \citep[e.g.][]{Brusa07}.
The estimator was computed for each survey and band
where a potential counterpart is present, and the highest $LR$ 
value was assigned to the counterpart set. We then ordered the
counterpart sets by decreasing $LR$ and divided them into
three broad groups: good ($LR>0.25$), fair ($0.05<LR<0.25$), and bad
($LR<0.25$). In addition,  as a cross-check we calculated the
simple probability of chance coincidence, according to
\citet{Downes86}.

We then assigned a preliminary rank, rejecting most of the
cases with bad scores. A primary single counterpart is either
a physical solitary association, a single non-bad association, or
exceptionally the best of the rejects that has been `recovered'. When
several candidates above the threshold exist, the one with the best
estimator is considered the primary counterpart, and all the others
 secondaries. The detailed procedure can be found in 
XXL paper XXVII.

\subsection{Optical photometry}
The BCS covers $\sim$50 deg$^2$, fully covering
XXL-S. Optical data in the griz bands were obtained with the Mosaic2
imager mounted on the Cerro Tololo 
Inter-American Observatory (CTIO) 4m Blanco telescope, reaching
10$\sigma$ point source depths of 23.9,
24.0, 23.6, and 22.1 mag (AB) in the four bands (for more
details see \citealt{Desai12}).
The DECam \citep{Flaugher15} mounted on the Blanco telescope also observed XXL-S (PI:
C. Lidman) in the griz bands ($4850-9000 \AA$). The limiting magnitudes 
in each band (defined as the third quartile of the
corresponding magnitude distribution) reach 25.73, 25.78, 25.6, and
24.87 mag (AB). More details of the observations
can be found in XXL paper VI and in \citet{Desai12,Desai15}.

\subsection{Optical spectroscopy}

A significant effort has been made to obtain spectroscopy of both
northern and southern XXL fields, either with large spectroscopic
surveys, for example SDSS, VIPERS \citep{Scodeggio16}, GAMA
\citep{Baldry18}, ESO Large Program \citep[][aka XXL paper XX]{Adami18}, 
or smaller scale spectroscopic observations, like the William
Herchel Telescope (WHT) spectroscopic follow-up program
\citep[][XXL paper XII]{Koulouridis16a}.
The southern field has been chosen for this study due to the
homogeneity of its spectroscopic follow-up data, which is based
uniquely on a number of observing runs with the multifibre 2dF+AAOmega
facility on the AAT;  instead, the
northern field  is based on a compilation of different surveys with
different instruments, limiting magnitudes, selection biases, and solid
angles. The first set of AAOmega runs occurred in 2013, and is
described in  \citet[][hereafter XXL paper XIV]{Lidman16}. The second
set of runs occurred in 2016 and is described in XXL paper XXVII. 
As noted in XXL paper XIV, priority was given to cluster galaxies over
AGN in the 2013 observations. In the 2016 set of observations, AGNs had the
highest priority. Hence, the coverage of the AGNs is, perhaps apart
from the smallest scales, uniform.

It  should also be remembered that there is always a lower fibre 
separation limit, which for the AAOmega case is $\sim 30^{''}-40^{''}$,
corresponding to $\sim 0.2 \;h^{-1}$ Mpc at the median redshift of our
sample. However, since any given region within the XXL area was
targeted multiple times with 2dF, the number of objects lost due to
potential fibre collisions was negligible.

The final XXL-S spectroscopic sample contains roughly $\sim 3740$ 
out of the $\sim 4100$ total X-ray point sources (a $\gtrsim$90\% completion) 
with $r$-band magnitude $\lesssim 21.8$, obtained during the two AAT
observing runs.
The fraction of sources that are stars is $\sim$10\%, and our final AGN
spectroscopic sample therefore consists of 3355 unique sources,  of
which 3106 are detected in the soft X-ray band and 1893 in the
hard.

\section{Methodology}
In order to quantify the low-order clustering of a distribution of sources, 
we use the two-point correlation function, $\xi(s)$, which describes
the excess probability over random of finding pairs of
sources within a range of redshift-space separations, $s$
\citep[e.g.][]{Peebles1980}. Therefore, when measuring $\xi$ directly
from redshift catalogues of sources, we 
unavoidably include the distorting effect of peculiar velocities 
since the comoving distance of a source is 
\begin{equation}
r=(s-{\bf v}_{p}\cdot{\bf r})/H_0 \;,
\end{equation}
where ${\bf v}_p\cdot {\bf r}$ 
is the component of the peculiar velocity of the 
 source along the line of sight. 
In order to avoid such effects the so-called projected correlation
function, $w_p(r_p)$, can be used to infer the spatial
  clustering \citep[e.g.][]{DavisPeeb1983}.
 
To this end,
we deconvolve the redshift-based distance of a source, $s$, in two
components, one parallel ($\pi$) and one perpendicular ($r_p$) to
the line of sight, i.e. $s=(r_p^2+\pi^2)^{1/2}$, and thus the redshift-space 
correlation function can be written as $\xi(s)=\xi(r_p, \pi)$.
Since redshift space distortions affect only the $\pi$
component, we can estimate the projected correlation function,
$w_p(r_p)$  (which is free of $z$-space distortions),
by integrating $\xi(r_p,\pi)$ along $\pi$:
\begin{equation}
w_p(r_p)=2\int_{0}^{\infty}\xi(r_p,\pi)\mathrm{d}\pi.
\end{equation}

Once we have estimated the projected correlation function, $w_p(r_p)$, we can
recover the real space correlation function since the two are related
according to \citep{DavisPeeb1983}
\begin{equation}\label{eq:wp}
 w_p(r_p)=2\int_{0}^{\infty}\xi(\sqrt{r_p^2+\pi^2}) {\rm d}\pi =2\int_{r_p}^{\infty}
 \frac{r\xi(r)\mathrm{d}r}{\sqrt{r^2-r_p^2}}\;.
 \end{equation}
Modelling $\xi(r)$ as a power law, $\xi(r)=\left(r/r_0\right)^{-\gamma}$,
we obtain
\begin{equation}\label{eq:wp_model}
w_p(r_p)=A_\gamma r_p \left(\frac{r_0}{r_p}\right)^{\gamma}
\end{equation}
with
\begin{equation}
A_\gamma=\Gamma\left(\frac{1}{2}\right)
\Gamma\left(\frac{\gamma-1}{2}\right)/\Gamma\left(\frac{\gamma}{2}\right),
\end{equation} 
where $\Gamma$ is the usual gamma function.

However, it should be noted that although Eq.~(\ref{eq:wp_model})
strictly holds for
$\pi_{\rm max}=\infty$, practically we always impose a cut-off
$\pi_{\rm max}$ (for reasons discussed in the next section). This
introduces an underestimation of the underlying correlation
function, which is an increasing function of separation $r_p$. 
For a power-law correlation function this underestimation is easily
inferred from Eq.~(\ref{eq:wp}) and is given by \citep[e.g.][]{Starikova11}
\begin{equation}
C_{\gamma}(r_p)=\frac{\int_0^{\pi_{\rm max}} (r_p^2+\pi^2)^{-\gamma/2} d\pi}
{\int_0^\infty (r_p^2+\pi^2)^{-\gamma/2} d\pi}\;.
\end{equation}
Thus, by taking into account the above statistical correction, and
under the assumption of the power-law correlation function,
we can recover the corrected spatial correlation function, $\xi(r_p)$, from
the fit to the measured $w_p(r_p)$ according to  
\begin{equation}\label{eq:corr}
\xi(r_p)=\frac{1}{A_\gamma C_{\gamma}(r_p)} \frac{w_p(r_p)}{r_p} \;,
\end{equation}
which also provides also
the value of $\gamma$. However, at large separations the correction factor increasingly
dominates over the signal, and thus it constitutes an unreliable correction
procedure. 

\subsection{Correlation function estimator}
As a first step we calculate $\xi(r_p,\pi)$ using the \citet{Landy93}
estimator \citep[for a discussion regarding different
  estimators see][]{Kerscher}
\begin{equation}\label{eq:xi}
1+\xi(r_p,\pi)=\frac{DD(r_p,\pi) - 2 DR(r_p,\pi) +
  RR(r_p,\pi)}{RR(r_p,\pi)}\;,
\end{equation} 
where $DD(r_p,\pi)$, $RR(r_p,\pi)$, and
$DR(r_p,\pi)$ are the number of data-data, random-random, and data-random pairs,
respectively. We then estimate the redshift-space correlation
function, $\xi(s)$, in the comoving redshift separation range $s\in [1,80] \;h^{-1}$
Mpc and the projected correlation function,
$w_p(r_p)$, in the comoving projected separation range $r_p\in[0.5,40] \;h^{-1}$ Mpc.
We note that large separations in the $\pi$ direction  mostly add noise to
the above estimator and therefore the integration is truncated for
separations larger than $\pi_{\rm max}$. The choice of  $\pi_{\rm
  max}$ is a compromise
between having an optimal signal-to-noise ratio for $\xi$ and reducing the
excess noise from high $\pi$ separations. The majority of studies in the
literature usually assume $\pi_{\rm max} \in [5, 30]~h^{-1}$ Mpc.

The correlation function uncertainty is estimated according to
\begin{equation}
\sigma_{w_p}=\sqrt{3} (1+w_p)/\sqrt{DD}\;,
\end{equation}
which corresponds to that expected by the bootstrap technique
\citep{Mo92}.
In this work, we bin the source pairs in logarithmic intervals of
 $\delta\log_{10}(r_p,\pi)\simeq 0.19$ and $\delta\log_{10}(s)\simeq 0.12$
for the $w_p(r_p)$ and $\xi(s)$ correlation functions, respectively.
Finally, we use a $\chi^2$ minimization 
procedure between data and the power-law
model for either type of the correlation function to derive the best-fit $r_0$
and {\rm$\gamma$} parameters. We carefully choose the range of separations  
in order to obtain the best  power-law fit to the data and we impose a lower
separation limit of $r_p\sim 1 \; h^{-1}$ Mpc to minimize non-linear effects.

\subsection{Random catalogue construction}
To estimate the spatial correlation function of a sample of sources
we need to construct a large mock comparison sample with a random spatial
distribution within the survey area, which also reproduces all the
systematic biases that are present in the source sample (i.e.
instrumental biases due to the point spread function variation,
vignetting, etc.). Also, special care has to be taken to reproduce any
biases that enter through the optical counterpart spectroscopic
observations strategy (for example, due to fibre collisions in
multifibre spectographs or due to the positioning of the slits on the
masks in multislit spectrographs, etc.).

To this end we  follow the random catalogue construction procedure
of \citet{Gilli05}, which is based on reshuffling
only the source redshifts, smoothing the corresponding redshift
distribution, while
keeping the angular coordinates unchanged, thus reproducing all the
previously discussed biases. 
In detail we assign random redshifts to the mock sample by smoothing
the source redshift distribution using a Gaussian
kernel with a smoothing length of $\sigma_z=0.125$.
This offers a compromise between scales
that are either too small and thus may reproduce the $z$-space
clustering, or too large and thus over-smooth the observed
redshift distribution, providing an unrealistically high clustering
pattern.  In Figure 1 we present the redshift distribution of the soft-
and hard-band AGN samples overplotted with the mean over 100 random
realizations, according to the above prescription.

\begin{figure}
\includegraphics[width=8.5cm]{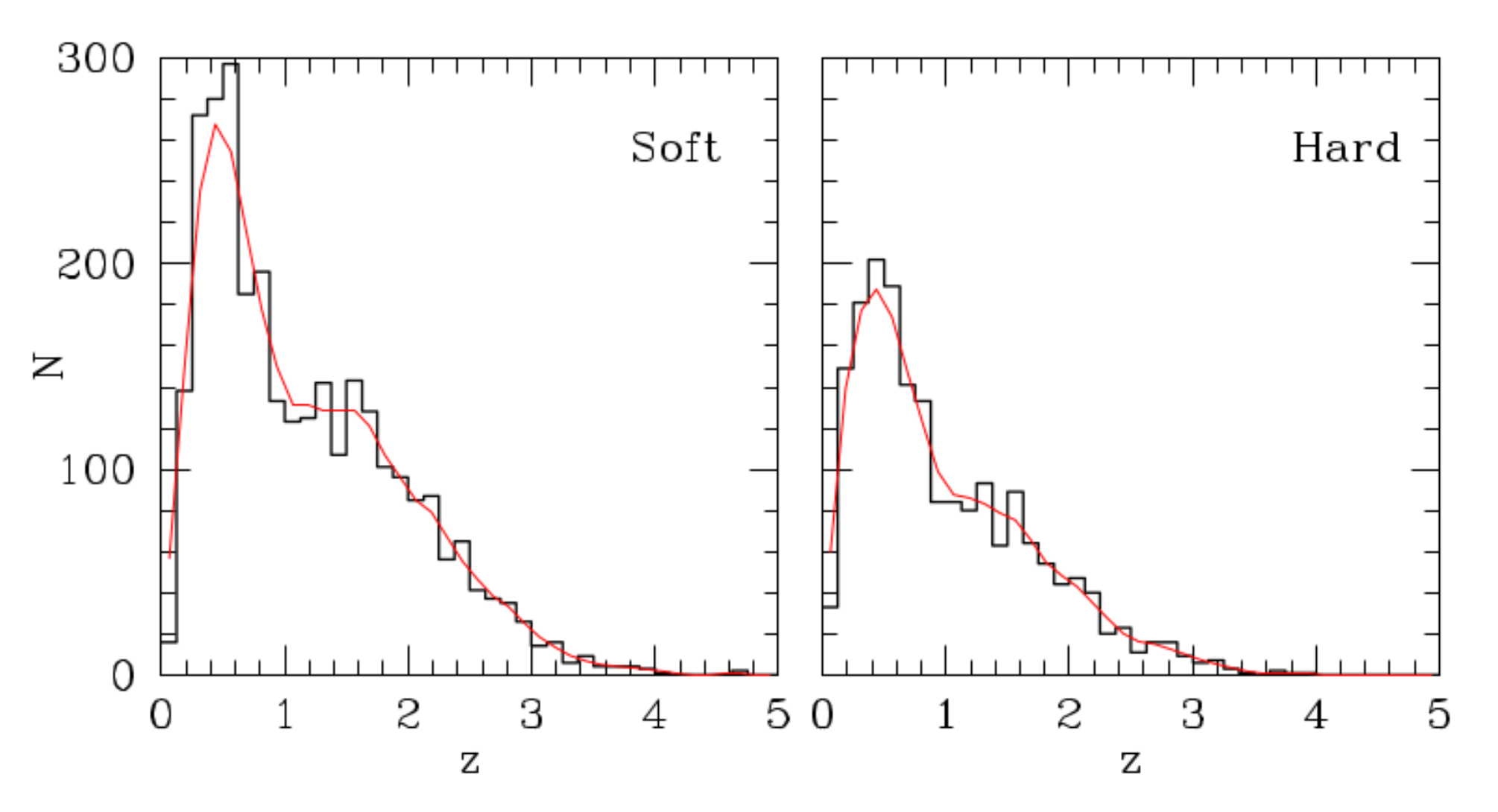}
\caption{Redshift distribution of the soft and hard XXL-S AGN
  samples. The red curve corresponds to the mean over 100 random realizations
  according to the prescription described in \citet{Gilli05} and using a Gaussian
with a smoothing length of $\sigma_z=0.125$.}
\end{figure}

\begin{figure}
\includegraphics[width=8.5cm]{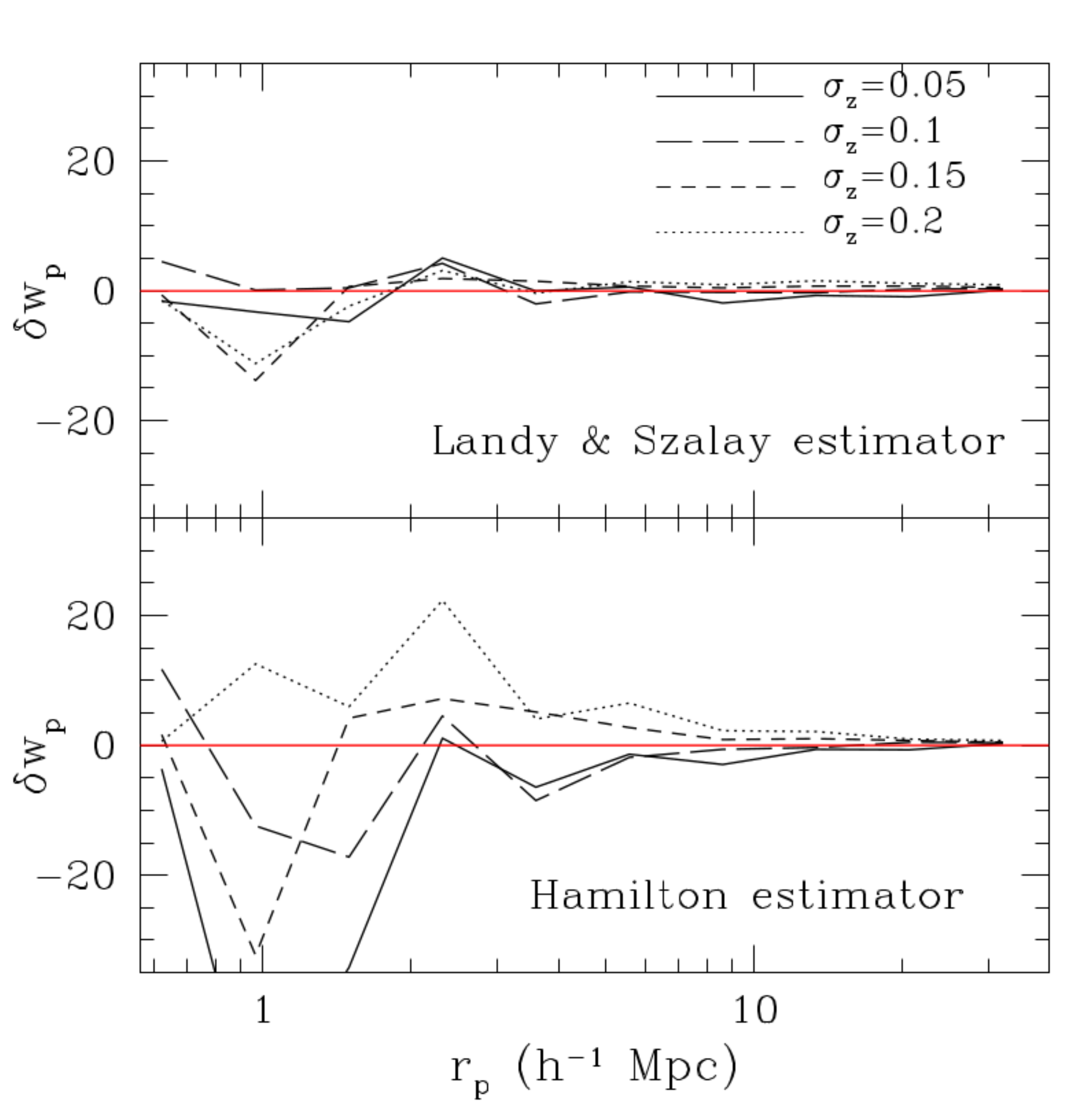}
\caption{Difference of the projected correlation function,
  $w_p(r_p)$, based on a random catalogues constructed with
  $\sigma_z=0.125$ with those based on the indicated values of
  $\sigma_z$. {\em Upper Panel:} Results based on the \citet{Landy93} 
 estimator of the 2-p correlation function.
{\em Lower Panel:} Results based on the \citet{Hamilton93} estimator of the 2-p correlation function.}
\end{figure}

We arrived at this value of $\sigma_z$ after investigating the effect of
different values on the resulting clustering
pattern. Interestingly, we found that values in the range 
$0.05\le \sigma_z \le 0.2$ provide very consistent clustering results
when using the \citet{Landy93} estimator of the 2-p correlation function, while
other estimators \citep[e.g.][]{Hamilton93} show a large scatter of
the clustering results. This is depicted in Figure 2 where we plot the $w_p(r)$
difference between the results based on $\sigma_z=0.125$ and those
indicated within the plot. It is evident that the Landy \& Szalay
estimator outperforms the corresponding  Hamilton estimator by
a large margin.

Finally, we note that in the following analysis we have limited our
samples to $z\le 3$, since the few higher redshift AGNs, sampling a
wide redshift range but extremely sparsely, would act mostly as noise
in the clustering analysis.

\section{Results and discussion}
We first present results based on the complete sample of sources,
spanning all redshifts, with
estimated $L_x\gtrsim 10^{41}$ erg s$^{-1}$ cm$^{-2}$, in order to
clearly exclude galaxies. We select the optimal $\pi_{max}$
cut-off by investigating the performance of the resulting clustering
parameters for different values of  $\pi_{max}$. We select as our
optimal value that for which the clustering parameters show stability.
In Figure 3 we present for both the soft- and hard-band sources 
the clustering length, $r_{p,0}$, as a function of
$\pi_{max}$ for the nominal slope $\gamma=1.8$. We see that for
$20\lesssim \pi_{max}/h^{-1} {\rm Mpc} \lesssim 40$  the value of
$r_{p,0}$ is quite stable and equals $\simeq 5.6$ and $\simeq 5.2$ $h^{-1}$ 
Mpc for the soft- and hard-band sources respectively. For the
remaining discussion and results we  use consistently
$\pi_{max}=20 \;h^{-1}$ Mpc.

\begin{figure}
\includegraphics[width=8.5cm]{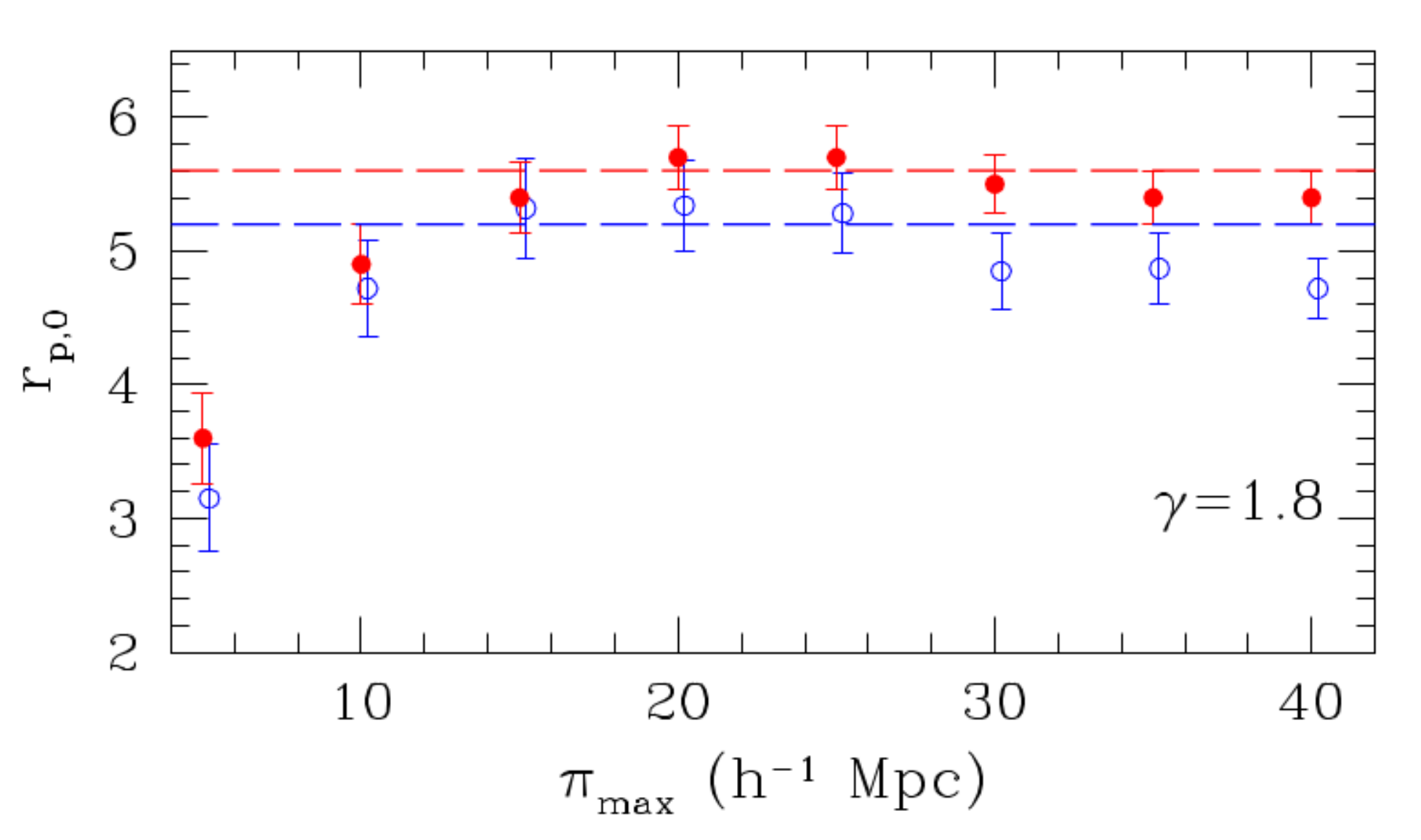}
\caption{ Dependence of the $w_p(r_p)$ clustering length on the
  cut-off $\pi_{max}$ value for the case of constant slope
  $\gamma=1.8$. The red filled points correspond to the soft-band
  results, while the blue open points to the corresponding hard-band
  results. The red and blue dashed lines correspond to the estimated final
  $r_{p,0}$ correlation lengths of the soft- and hard-band sources,
  respectively.}

\end{figure}

The derived $w_p(r_p)$ correlation function is shown in
Figure 4 for both soft- and hard-band sources, while the parameters of
the power-law fit, within $1\; h^{-1} {\rm Mpc} \lesssim r_p \lesssim
25 \; h^{-1}$ Mpc, are shown in Table 1 for both the projected
correlation function, $w_p(r_p),$ and the inferred spatial correlation function
(Eq. \ref{eq:corr}).

A first important result, indicated both by Figure 4 and Table 1, is the
clustering consistency between the soft- and hard-band sources,
although
there seems to be a slight but rather insignificant enchancement of the
soft-band clustering with respect to that of the hard-band, indicated
also in the inset panel by the confidence levels of the fitted
correlation function parameters. The inferred spatial correlation
length is $r_{p,0}=7.00\pm0.34$ and $6.42\pm0.42$ $h^{-1}$ Mpc for the
soft- and hard-band sources, respectively, while for both the slope is
$\gamma=1.44\pm 0.10$. We also note  the median
redshift difference among the two bands.

A second important result is that the inferred clustering is in good
agreement with Chandra results, i.e. with the
\citet{Koutoulidis13} clustering analysis of a large compilation of 1466
Chandra 0.5--8 keV sources, which have a median redshift of
$\bar{z}\simeq 0.98$, similar to that of our sample, and which
provided $r_0=7.2 (\pm 0.6) \; h^{-1}$ Mpc and $\gamma=1.48 (\pm
0.12)$, and  with the \citet{Starikova11} results of the
Bootes field (based on the 0.5--2 keV band sources).

\begin{table}
\caption{Clustering results based on a power-law model fit
 within $1<r_p<20 \; h^{-1}$ Mpc, using the whole sample of soft- and
  hard-band X-ray AGN sources. The clustering length units are $h^{-1}$
  Mpc. The results correspond to $\pi_{\rm max}=20 h^{-1}$ Mpc.}
\tabcolsep 8pt
\begin{tabular}{l c c c c } \hline 
            & band & $\gamma$        &  $r_0$        & $r_0$ ($\gamma=1.8$)  \\  \hline
$w_p(r_p)$  & soft & 1.79$\pm0.02$  & 5.57$\pm0.12$ & $5.62\pm0.13$ \\
$w_p(r_p)$  & hard & 1.81$\pm0.02$  & 5.29$\pm0.16$ & $5.25\pm0.16$
  \\ \\
$\xi(r_p)$  & soft & 1.44$\pm 0.08$ & 7.00$\pm0.34$ & $7.52\pm0.30$ \\ 
$\xi(r_p)$  & hard & 1.44$\pm 0.10$ & 6.42$\pm0.42$ & $6.98\pm0.37$
  \\ \\
$\xi(s)$  & soft & 1.32$\pm 0.17$ & 6.50$\pm0.72$ & $6.79\pm0.58$ \\ 
$\xi(s)$  & hard & 1.41$\pm 0.25$ & 6.26$\pm0.89$ & $6.49\pm0.75$ \\ 
\hline

\end{tabular}
\end{table}

\begin{figure}
\includegraphics[width=8.1cm]{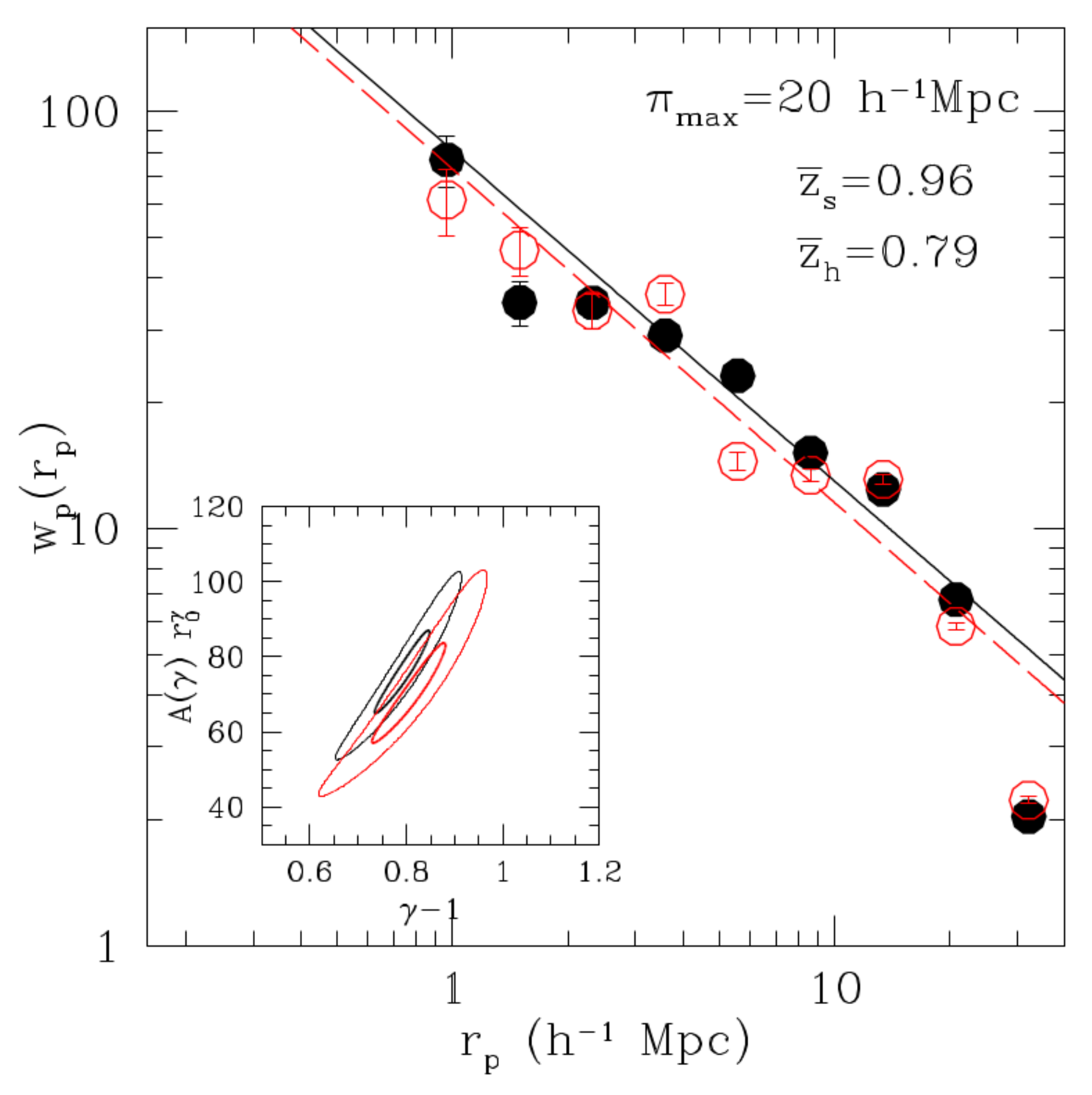}
\caption{ Projected correlation function of the soft
  (filled points) and hard X-ray AGNs (open red points) for the whole
  XXL-S AGN sample and for $r_p\gtrsim 1 \; h^{-1}$ Mpc. 
In the inset panel we only show the 1$\sigma$ and
3$\sigma$ confidence contour levels of the corresponding fitted
correlation function parameters in order to avoid crowding.}
\end{figure}

In order to visualize the recovery of the true spatial correlation
function via Eq. \ref{eq:corr}, we plot in Figure 5 the inferred
correlation function $\xi(r_p)$ together with the redshift-space
correlation function, $\xi(s)$ (separately for the soft- and hard-band XXL-S
sources) which should be boosted up by redshift-space distortions. 
Evidently, however, there is an excellent consistency between the two, with both
the amplitude and the slope of the power-law fit being in
agreement, as  can be seen in the insets  and in Table 1. We
infer that redshift-space distortions do not significantly affect our
sample. Furthermore, the inferred spatial clustering lengths
correspond to bias values of $b\simeq 2.17\pm 0.10$ and $b\simeq
1.87\pm 0.11$ for the soft- and hard-bands at $\bar{z}=0.96$ and 0.79,
respectively. These bias values result from \citep{Peebles1980,Peebles1993}
\begin{equation}
b(z)=\left(\frac{r_0}{8}\right)^{\frac{\gamma}{2}} J_2^{\frac{1}{2}}
\left(\frac{\sigma_8D(z)}{D(0)}\right)^{-1} \;,
\end{equation} 
where $D(z)$ is the $\Lambda$CDM fluctuations growing mode, $\sigma_8=0.81$
is the normalization of the power spectrum 
(consistent with that of the joint Planck analysis of the \citet{Ade16}), and
$J_2(\gamma)=72/[(3-\gamma) (4-\gamma) (6-\gamma) 2^{\gamma}]$.


\begin{figure}
\includegraphics[width=8.6cm]{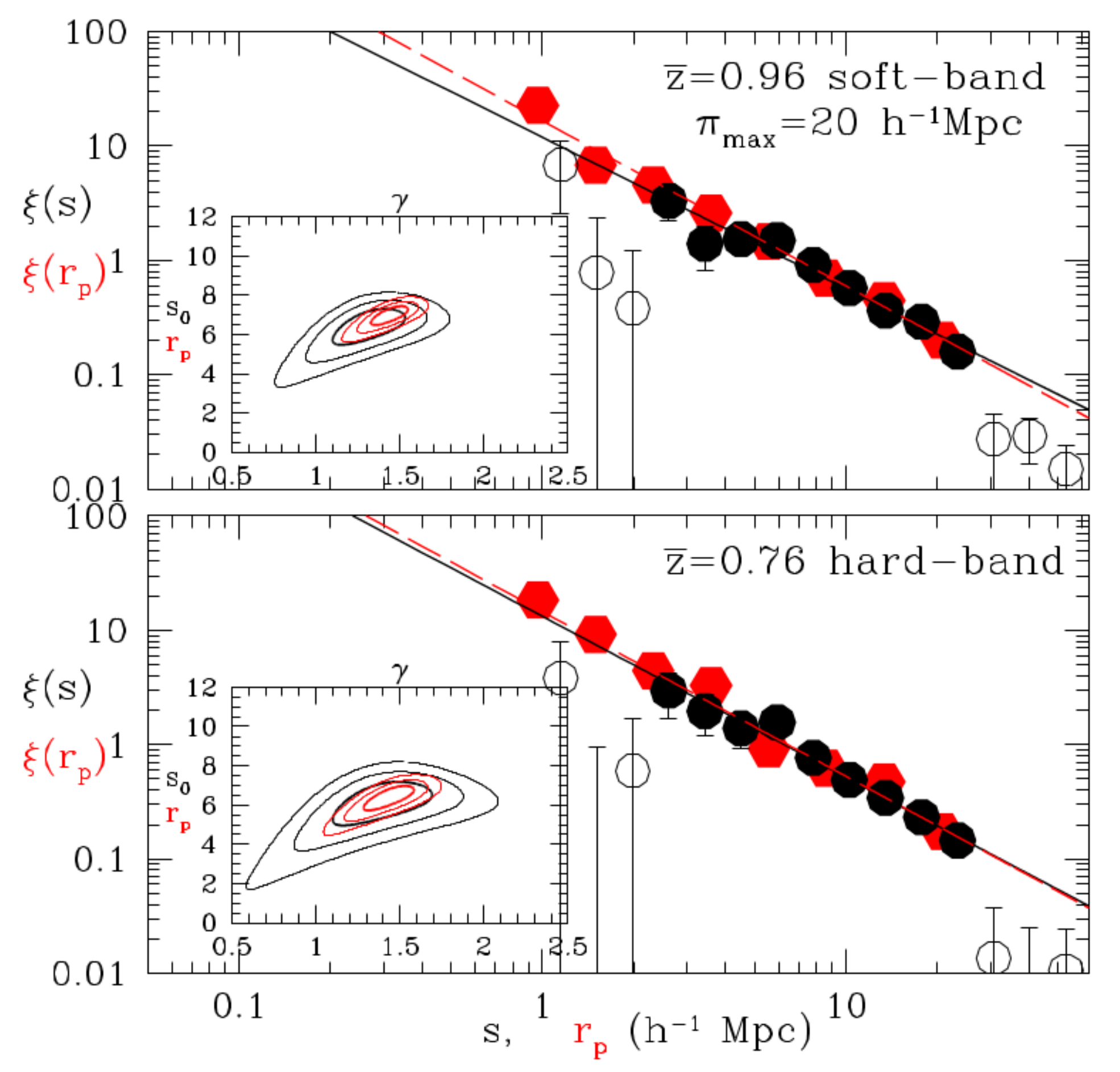}
\caption{ Redshift-space correlation function, $\xi(s)$ (filled
  circular points), and the inferred real 3D correlation function, $\xi(r_p)$
   (red pentagons), for the soft-
  (upper panel) and hard-band XXL-S AGNs (lower panel). 
The black line corresponds to the power-law fit to $\xi(s),$ and the
red line to $\xi(r_p)$. The empty circular points signify the small
and large separation $\xi(s)$ range that are not used in the
power-law fit. In the insets we show the 1, 2, and
3$\sigma$ confidence contour levels of the corresponding fitted
correlation function parameters.}
\end{figure}

We now attempt to investigate the redshift evolution of the AGN
correlation function for both the soft- and hard-band sources. 
To this end we determine the clustering pattern of the following:
\begin{itemize}
\item the dominant population of XXL-S sources, i.e. those that
populate the redshift range around its mode value ($0.3<z<1.1$), 
which consists of 1375
and 932 soft- and hard-band sources, respectively, with a median
redshift of the samples $\bar{z}=0.59$; 
\item the higher redshift regime, i.e. $1.1<z<3$, which consists
of 1291 soft and 680 hard band sources, respectively, with
corresponding median redshifts: $\bar{z}=1.73$ and 1.63.
\end{itemize}

The resulting spatial clustering power-law results for the $0.3<z<1.1$
redshift bin are $r_0=7.10\pm 0.42 \; h^{-1}$ Mpc and $\gamma=1.45\pm 0.09$ 
for the soft-band and $r_0=5.52\pm 0.54\; h^{-1}$ Mpc  and
$\gamma=1.42\pm 0.12$ for the hard-band, respectively. 
The 1, 2, and $3\sigma$ confidence contour levels of the fitted two
parameter power-law correlation function model can be seen in Figure 6.
The difference between the soft- and hard-band results is now more pronounced,
providing evidence for the reality of the difference.
These  clustering lengths corresponds to bias values of $b\simeq
1.85\pm 0.10$ and $b\simeq 1.56\pm 0.13$, respectively. 
\begin{figure}
\includegraphics[width=8.9cm]{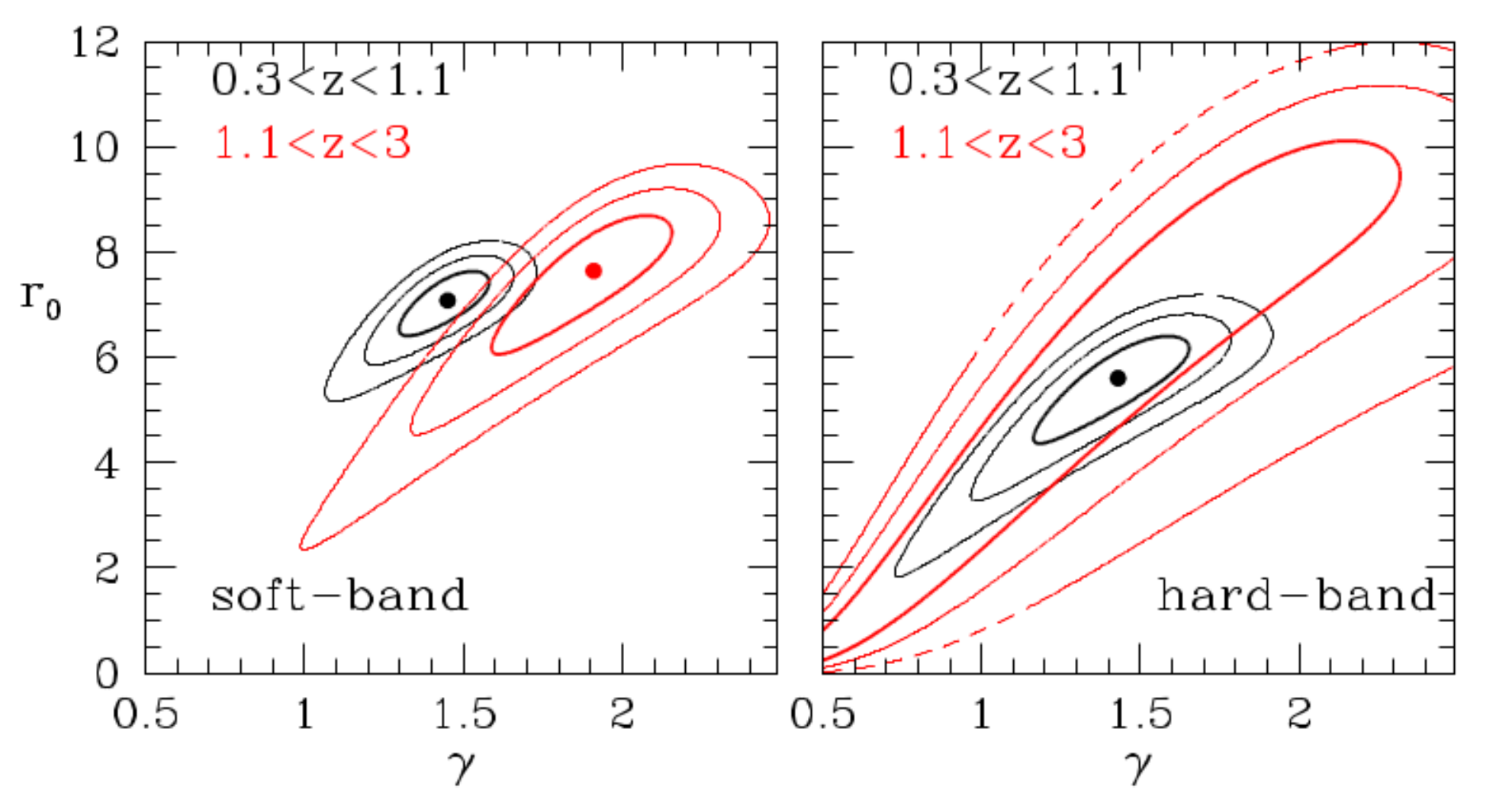}
\caption{Confidence contour levels (1, 2, and $3\sigma$)  of the
$\chi^2$ fitted two-parameter power-law correlation function model for two redshift
ranges, and separately for the soft- and hard-band sources.}
\end{figure}
The results appear to contradict the findings of \citet{Elyiv12},
based on the XMM-LSS, who found that the amplitude of the correlation
function is higher in the hard
band than in the soft band. A possible explanation is that
they do not derive the correlation length  directly (due to the absence of
spectroscopic redshift information), but use the more uncertain
Limbers inversion of the angular clustering pattern. 

Analysing the soft-band AGN sample in the higher redshift bin, we find
results with a large scatter depending on the separation range
used to fit the power-law model to the correlation function;
it ranges from $\sim 6$ to 8
$h^{-1}$ Mpc and the slope $\gamma$ from $\sim$ 1.2 to $\sim$2. 
Using the separation range $r_p\in[3,35]\; h^{-1}$ Mpc, in order to
avoid the large scatter at smaller separations and to enhance our signal,
we find $r_0= 7.64\pm 0.7\; h^{-1}$ Mpc and $\gamma=1.91\pm0.16$,
while for a fixed $\gamma=1.8$ we obtain $r_0\simeq 7.2\pm 1.0\; h^{-1}$ Mpc.
The measured clustering corresponds to a linear bias factor of
$b=3.68 \pm 0.46$ at $\bar{z}=1.73$, where the quoted uncertainty
also takes into account  the $r_0$ scatter due to the separation range
used in the power-law fit.

\begin{figure}
\includegraphics[width=8.1cm]{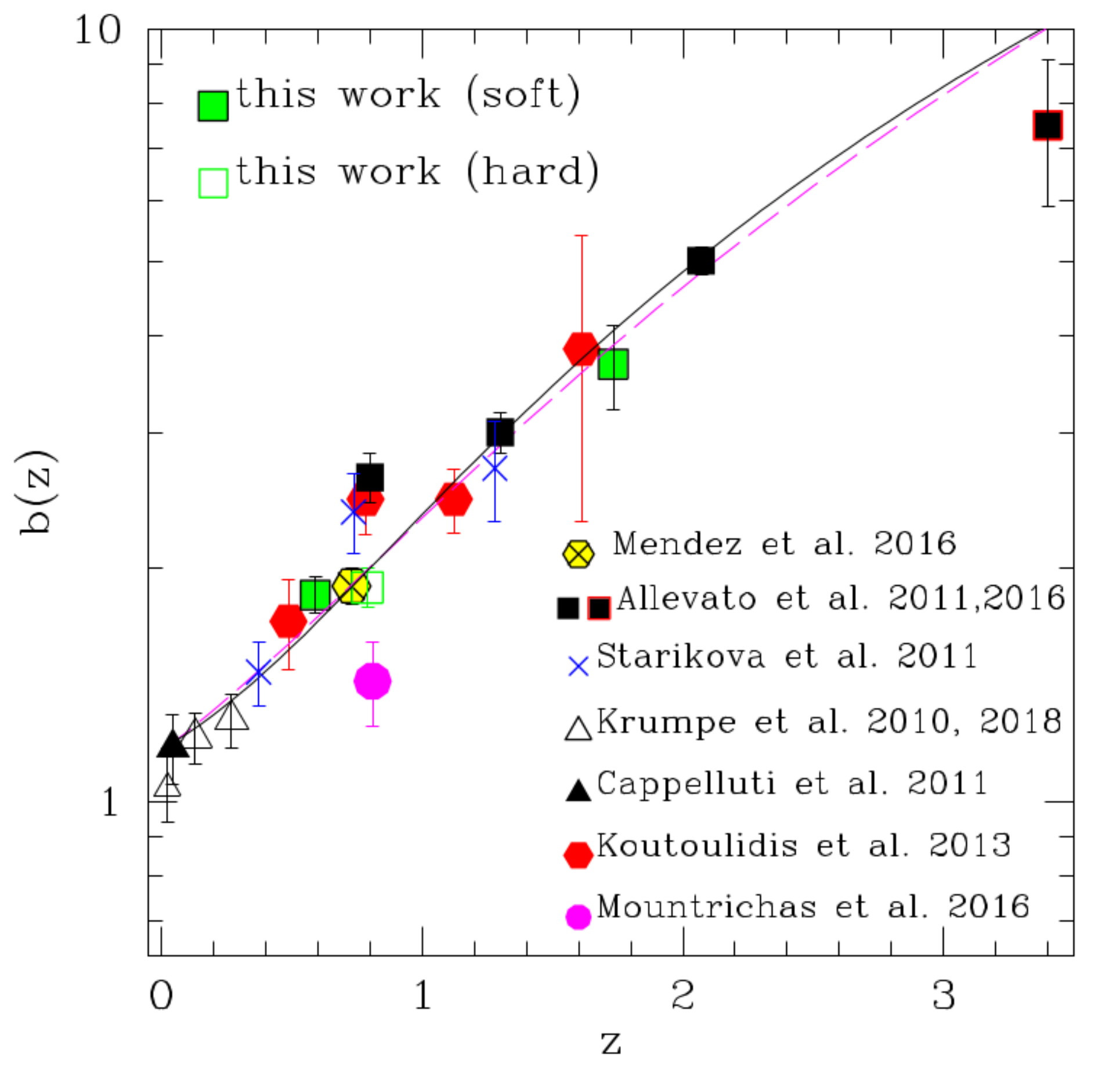}
\caption{Redshift evolution of the X-ray AGN linear bias factor,
  as derived from a number of recent studies indicated in the
  plot. The bias values shown correspond to a flat $\Lambda$CDM
  cosmological model with $\Omega_m=0.3$ and $\sigma_8=0.81$.
  The black continuous curve 
  corresponds to the bias evolution model of \citet{Basilakos08} for a
  DM host halo mass of $M_h=10^{13.04} h^{-1} M_{\odot}$, while the magenta
  dashed line corresponds to the model of \citet{TRK} for
  $M_h=10^{12.96} h^{-1} M_{\odot}$.}
\end{figure}

In order to compare our results with those of a large number of recent
determinations of the X-ray AGN bias, we present in Figure 7 the
corresponding bias values as a function of redshift. The results of
different studies are all translated to the {\em Planck 2015}
cosmology ($\Omega_m=0.3$ and $\sigma_8=0.81$) and are 
indicated by different symbols (as indicated within
the figure). The filled green symbols correspond to our current results based 
on the soft band and the two separate redshift bins, while the
open green square corresponds to the overall hard-band results (since
we were unable to clearly detect a clustering signal at the high
redshift range). 

We have fitted  two different bias
evolution models to the data, the \citet{Basilakos08} and the
\citet{TRK} models,
using a $\chi^2$ minimization procedure and we derive the DM host halo
mass \citep[details and comparisons of the different bias models can
  be found in][]{Papageorgiou12,Papageorgiou17}. 
We find $\log_{10}M_h/[h^{-1} M_{\odot}]=13.04\pm 0.06$ and 
$12.98\pm 0.07$ for the two models, respectively. The corresponding
fits can be seen in Figure 6 as the continous black and broken magenta curves.
We note that in the above fit we have excluded the offset results 
at $z\sim 0.75$ and by doing so we obtain an
excellent fit with a reduced $\chi^2$/d.f.$=0.93$. Using all the bias data
we obtain the same halo mass, but naturally with a worse reduced
$\chi^2$. 

It is evident that the current results are in agreement with the
general trend provided by all previous studies, and consistent with
X-ray AGN being hosted by $\sim 10^{13} \; h^{-1} M_{\odot}$ DM halos.
The general trend of a significantly higher bias and corresponding
host DM halo mass of the X-ray selected AGN with respect to
optically selected AGN \citep[e.g.][]{Croom05,Ross09}, which appear
to inhabit DM halos with $\log_{10}M_h/[h^{-1} M_{\odot}]\simeq
12.50\pm 0.05$ \citep[see e.g.][and references
  therein]{Koutoulidis13}, and indicates a different fuelling mechanism
for the two populations of AGN.

\section{Conclusions}
We have analysed the clustering pattern of a homogeneous X-ray AGN
spectroscopic sample of the XMM southern field, with a $\sim 90\%$
redshift completeness down to an optical counterpart $r$-band magnitude
of 21.8. The sample constitutes the largest spectroscopic sample of
X-ray AGN, with
a flux limit of $\sim 2.2 \times 10^{-15}$ and $\sim 1.4 \times
10^{-14}$ erg s$^{-1}$ cm$^{2}$ (corresponding to 50\% of the area
curve for the
soft band and hard band, respectively), covering a coherent area of
$\sim 25$ deg$^2$.

Our main results are as follows:
\begin{itemize}
\item The clustering of the X-ray AGN, detected in either soft or hard
  bands, are well represented by the usual power law in the separation
  range of $1\lesssim r_p\lesssim 25 \; h^{-1}$ Mpc, with the inferred
  spatial clustering length being $r_0\simeq 7 (\pm 0.3)$ and 6.4$(\pm 0.4)$
  $h^{-1}$ Mpc for the soft and hard-band detected sources,
  respectively. The slope
  for both cases is $\gamma=1.44 (\pm 0.1)$. The corresponding clustering
  lengths for the nominal slope $\gamma=1.8$ are $r_0\simeq 7.5 (\pm 0.3)$ and
  $\simeq 7.0 (\pm 0.4)$ $h^{-1}$ Mpc. These results are in good
  agreement with the analysis of a joint Chandra sample of 1466
  sources, having a similar median redshift with our sample,
 by \citet{Koutoulidis13} who find $r_0\simeq 7.2 (\pm 0.6) \; h^{-1}$
Mpc and $\gamma=1.48 (\pm 0.12)$.
\item The weak excess clustering of the soft sources with respect to those
  detected in the hard band becomes more pronounced and significant
  if we limit our analysis around the mode of the redshift
  distribution (i.e. $0.3<z<1.1$), in which case we find the same
  slope of the power-law, but the clustering lengths become $\sim 7.1
  (\pm 0.4)$  and $\sim 5.5 (\pm 0.5)$ $h^{-1}$ Mpc for the soft- and hard-band detected sources,
  respectively. This is in
  disagreement with the angular clustering analysis of \citet{Elyiv12}.
\item The derived linear bias factor at the median redshift of the
  sample and in two separate redshift bins corresponds to the expectation of
  host dark matter halos with a mass $M_h\simeq 10^{13} h^{-1} \;
  M_{\odot}$, in agreement with most recent analysis of local or
  distant samples of X-ray AGN \citep[e.g.][]{Allevato16,Mendez16,Krumpe18}. 
\end{itemize}

\begin{acknowledgements}
The Saclay group acknowledges long-term support from
the Centre National d'Etudes Spatiales (CNES). EK thanks CNES and CNRS
for their support of post-doctoral research.
XXL is an international project based around an XMM Very Large
Programme surveying two 25 deg$^2$ extragalactic fields at a depth of
$\sim 5\times 10^{-15}$ erg cm$^{-2}$ s$^{-1}$ in the [0.5--2] keV band
for point-like
sources. The XXL website is http://irfu.cea.fr/xxl. Multi-band
information and spectroscopic follow-up of the X-ray sources are
obtained through a number of survey programmes, summarised at
http://xxlmultiwave.pbworks.com/.
\end{acknowledgements}

\bibliographystyle{aa}
\bibliography{ms}

\begin{thebibliography}{65}
\expandafter\ifx\csname natexlab\endcsname\relax\def\natexlab#1{#1}\fi

\bibitem[{{Adami} {et~al.}(2018){Adami}, {Giles}, {Koulouridis}, {Balogh},
  {Bower}, {Smail}, {Ziegler}, {Davies}, {Gaztelu}, \& {Fritz}}]{Adami18}
{Adami}, C., {Giles}, P., {Koulouridis}, E., {et~al.} 2018, XXL paper XX,
  accepted in A\&A

\bibitem[{{Alexander} \& {Hickox}(2012)}]{Alexander12}
{Alexander}, D.~M. \& {Hickox}, R.~C. 2012, \nar, 56, 93

\bibitem[{{Allevato} {et~al.}(2016){Allevato}, {Civano}, {Finoguenov},
  {Marchesi}, {Shankar}, {Zamorani}, {Hasinger}, {Salvato}, {Miyaji}, {Gilli},
  {Cappelluti}, {Brusa}, {Suh}, {Lanzuisi}, {Trakhtenbrot}, {Griffiths},
  {Vignali}, {Schawinski}, \& {Karim}}]{Allevato16}
{Allevato}, V., {Civano}, F., {Finoguenov}, A., {et~al.} 2016, \apj, 832, 70

\bibitem[{{Allevato} {et~al.}(2011){Allevato}, {Finoguenov}, {Cappelluti},
  {Miyaji}, {Hasinger}, {Salvato}, {Brusa}, {Gilli}, {Zamorani}, {Shankar},
  {James}, {McCracken}, {Bongiorno}, {Merloni}, {Peacock}, {Silverman}, \&
  {Comastri}}]{Allevato11}
{Allevato}, V., {Finoguenov}, A., {Cappelluti}, N., {et~al.} 2011, ApJ, 736, 99

\bibitem[{{Baldry} {et~al.}(2018){Baldry}, {Liske}, {Brown}, {Robotham},
  {Driver}, {Dunne}, {Alpaslan}, {Brough}, {Cluver}, {Eardley}, {Farrow},
  {Heymans}, {Hildebrandt}, {Hopkins}, {Kelvin}, {Loveday}, {Moffett},
  {Norberg}, {Owers}, {Taylor}, {Wright}, {Bamford}, {Bland-Hawthorn},
  {Bourne}, {Bremer}, {Colless}, {Conselice}, {Croom}, {Davies}, {Foster},
  {Grootes}, {Holwerda}, {Jones}, {Kafle}, {Kuijken}, {Lara-Lopez},
  {L{\'o}pez-S{\'a}nchez}, {Meyer}, {Phillipps}, {Sutherland}, {van Kampen}, \&
  {Wilkins}}]{Baldry18}
{Baldry}, I.~K., {Liske}, J., {Brown}, M.~J.~I., {et~al.} 2018, \mnras, 474,
  3875

\bibitem[{{Basilakos} {et~al.}(2008){Basilakos}, {Plionis}, \&
  {Ragone-Figueroa}}]{Basilakos08}
{Basilakos}, S., {Plionis}, M., \& {Ragone-Figueroa}, C. 2008, \apj, 678, 627

\bibitem[{{Bertin} \& {Arnouts}(1996)}]{bertin96}
{Bertin}, E. \& {Arnouts}, S. 1996, \aaps, 117, 393

\bibitem[{{Bournaud} {et~al.}(2011){Bournaud}, {Dekel}, {Teyssier}, {Cacciato},
  {Daddi}, {Juneau}, \& {Shankar}}]{Bournaud11}
{Bournaud}, F., {Dekel}, A., {Teyssier}, R., {et~al.} 2011, ApJ, 741, L33

\bibitem[{{Brusa} {et~al.}(2007){Brusa}, {Zamorani}, {Comastri}, {Hasinger},
  {Cappelluti}, {Civano}, {Finoguenov}, {Mainieri}, {Salvato}, {Vignali},
  {Elvis}, {Fiore}, {Gilli}, {Impey}, {Lilly}, {Mignoli}, {Silverman}, {Trump},
  {Urry}, {Bender}, {Capak}, {Huchra}, {Kneib}, {Koekemoer}, {Leauthaud},
  {Lehmann}, {Massey}, {Matute}, {McCarthy}, {McCracken}, {Rhodes}, {Scoville},
  {Taniguchi}, \& {Thompson}}]{Brusa07}
{Brusa}, M., {Zamorani}, G., {Comastri}, A., {et~al.} 2007, \apjs, 172, 353

\bibitem[{{Cappelluti} {et~al.}(2010){Cappelluti}, {Ajello}, {Burlon},
  {Krumpe}, {Miyaji}, {Bonoli}, \& {Greiner}}]{Cappelluti10}
{Cappelluti}, N., {Ajello}, M., {Burlon}, D., {et~al.} 2010, ApJ, 716, L209

\bibitem[{{Cash}(1979)}]{cash79}
{Cash}, W. 1979, \apj, 228, 939

\bibitem[{{Chiappetti} {et~al.}(2018){Chiappetti}, {Fotopoulou}, {Lidman},
  {Balogh}, {Bower}, {Smail}, {Ziegler}, {Davies}, {Gaztelu}, \&
  {Fritz}}]{Chiappetti18}
{Chiappetti}, L., {Fotopoulou}, S., {Lidman}, C., {et~al.} 2018, XXL paper XX,
  accepted in A\&A

\bibitem[{{Coil} {et~al.}(2009){Coil}, {Georgakakis}, {Newman}, {Cooper},
  {Croton}, {Davis}, {Koo}, {Laird}, {Nandra}, {Weiner}, {Willmer}, \&
  {Yan}}]{Coil09}
{Coil}, A.~L., {Georgakakis}, A., {Newman}, J.~A., {et~al.} 2009, \apj, 701,
  1484

\bibitem[{{Coil} {et~al.}(2007){Coil}, {Hennawi}, {Newman}, {Cooper}, \&
  {Davis}}]{Coil07}
{Coil}, A.~L., {Hennawi}, J.~F., {Newman}, J.~A., {Cooper}, M.~C., \& {Davis},
  M. 2007, \apj, 654, 115

\bibitem[{{Croom} {et~al.}(2005){Croom}, {Boyle}, {Shanks}, {Smith}, {Miller},
  {Outram}, {Loaring}, {Hoyle}, \& {da {\^A}ngela}}]{Croom05}
{Croom}, S.~M., {Boyle}, B.~J., {Shanks}, T., {et~al.} 2005, MNRAS, 356, 415

\bibitem[{{Davis} \& {Peebles}(1983)}]{DavisPeeb1983}
{Davis}, M. \& {Peebles}, P.~J.~E. 1983, ApJ, 267, 465

\bibitem[{{Desai} {et~al.}(2012){Desai}, {Armstrong}, {Mohr}, {Semler}, {Liu},
  {Bertin}, {Allam}, {Barkhouse}, {Bazin}, {Buckley-Geer}, {Cooper}, {Hansen},
  {High}, {Lin}, {Lin}, {Ngeow}, {Rest}, {Song}, {Tucker}, \&
  {Zenteno}}]{Desai12}
{Desai}, S., {Armstrong}, R., {Mohr}, J.~J., {et~al.} 2012, \apj, 757, 83

\bibitem[{{Desai} {et~al.}(2015){Desai}, {Mohr}, {Henderson}, {K{\"u}mmel},
  {Paech}, \& {Wetzstein}}]{Desai15}
{Desai}, S., {Mohr}, J.~J., {Henderson}, R., {et~al.} 2015, Journal of
  Instrumentation, 10, C06014

\bibitem[{{Di Matteo} {et~al.}(2005){Di Matteo}, {Springel}, \&
  {Hernquist}}]{DiMatteo05}
{Di Matteo}, T., {Springel}, V., \& {Hernquist}, L. 2005, in Growing Black
  Holes: Accretion in a Cosmological Context, ed. A.~{Merloni}, S.~{Nayakshin},
  \& R.~A. {Sunyaev}, 340--345

\bibitem[{{Downes} {et~al.}(1986){Downes}, {Peacock}, {Savage}, \&
  {Carrie}}]{Downes86}
{Downes}, A.~J.~B., {Peacock}, J.~A., {Savage}, A., \& {Carrie}, D.~R. 1986,
  \mnras, 218, 31

\bibitem[{{Elyiv} {et~al.}(2012){Elyiv}, {Clerc}, {Plionis}, {Surdej},
  {Pierre}, {Basilakos}, {Chiappetti}, {Gandhi}, {Gosset}, {Melnyk}, \&
  {Pacaud}}]{Elyiv12}
{Elyiv}, A., {Clerc}, N., {Plionis}, M., {et~al.} 2012, AAP, 537, A131

\bibitem[{{Fanidakis} {et~al.}(2012){Fanidakis}, {Baugh}, {Benson}, {Bower},
  {Cole}, {Done}, {Frenk}, {Hickox}, {Lacey}, \& {Del P.~Lagos}}]{Fanidakis12}
{Fanidakis}, N., {Baugh}, C.~M., {Benson}, A.~J., {et~al.} 2012, MNRAS, 419,
  2797

\bibitem[{{Fanidakis} {et~al.}(2013){Fanidakis}, {Georgakakis}, {Mountrichas},
  {Krumpe}, {Baugh}, {Lacey}, {Frenk}, {Miyaji}, \& {Benson}}]{Fanidakis13}
{Fanidakis}, N., {Georgakakis}, A., {Mountrichas}, G., {et~al.} 2013, \mnras,
  435, 679

\bibitem[{{Flaugher} {et~al.}(2015){Flaugher}, {Diehl}, {Honscheid}, {Abbott},
  {Alvarez}, {Angstadt}, {Annis}, {Antonik}, {Ballester}, {Beaufore},
  {Bernstein}, {Bernstein}, {Bigelow}, {Bonati}, {Boprie}, {Brooks},
  {Buckley-Geer}, {Campa}, {Cardiel-Sas}, {Castander}, {Castilla}, {Cease},
  {Cela-Ruiz}, {Chappa}, {Chi}, {Cooper}, {da Costa}, {Dede}, {Derylo},
  {DePoy}, {de Vicente}, {Doel}, {Drlica-Wagner}, {Eiting}, {Elliott}, {Emes},
  {Estrada}, {Fausti Neto}, {Finley}, {Flores}, {Frieman}, {Gerdes},
  {Gladders}, {Gregory}, {Gutierrez}, {Hao}, {Holland}, {Holm}, {Huffman},
  {Jackson}, {James}, {Jonas}, {Karcher}, {Karliner}, {Kent}, {Kessler},
  {Kozlovsky}, {Kron}, {Kubik}, {Kuehn}, {Kuhlmann}, {Kuk}, {Lahav}, {Lathrop},
  {Lee}, {Levi}, {Lewis}, {Li}, {Mandrichenko}, {Marshall}, {Martinez},
  {Merritt}, {Miquel}, {Mu{\~n}oz}, {Neilsen}, {Nichol}, {Nord}, {Ogando},
  {Olsen}, {Palaio}, {Patton}, {Peoples}, {Plazas}, {Rauch}, {Reil}, {Rheault},
  {Roe}, {Rogers}, {Roodman}, {Sanchez}, {Scarpine}, {Schindler}, {Schmidt},
  {Schmitt}, {Schubnell}, {Schultz}, {Schurter}, {Scott}, {Serrano}, {Shaw},
  {Smith}, {Soares-Santos}, {Stefanik}, {Stuermer}, {Suchyta}, {Sypniewski},
  {Tarle}, {Thaler}, {Tighe}, {Tran}, {Tucker}, {Walker}, {Wang}, {Watson},
  {Weaverdyck}, {Wester}, {Woods}, {Yanny}, \& {DES
  Collaboration}}]{Flaugher15}
{Flaugher}, B., {Diehl}, H.~T., {Honscheid}, K., {et~al.} 2015, \aj, 150, 150

\bibitem[{{Fotopoulou} {et~al.}(2016){Fotopoulou}, {Buchner},
  {Georgantopoulos}, {Hasinger}, {Salvato}, {Georgakakis}, {Cappelluti},
  {Ranalli}, {Hsu}, {Brusa}, {Comastri}, {Miyaji}, {Nandra}, {Aird}, \&
  {Paltani}}]{Fotopoulou16}
{Fotopoulou}, S., {Buchner}, J., {Georgantopoulos}, I., {et~al.} 2016, \aap,
  587, A142

\bibitem[{{Gilli} {et~al.}(2005){Gilli}, {Daddi}, {Zamorani}, {Tozzi},
  {Borgani}, {Bergeron}, {Giacconi}, {Hasinger}, \& {Mainieri}}]{Gilli05}
{Gilli}, R., {Daddi}, E., {Zamorani}, G., {et~al.} 2005, A\&A, 430, 811

\bibitem[{{Gilli} {et~al.}(2009){Gilli}, {Zamorani}, {Miyaji}, {Silverman},
  {Brusa}, {Mainieri}, {Cappelluti}, {Daddi}, {Porciani}, {Pozzetti}, {Civano},
  \& {Comastri}}]{Gilli09}
{Gilli}, R., {Zamorani}, G., {Miyaji}, T., {et~al.} 2009, A\&A, 494, 33

\bibitem[{{G{\"u}ltekin} {et~al.}(2009){G{\"u}ltekin}, {Richstone}, {Gebhardt},
  {Lauer}, {Tremaine}, {Aller}, {Bender}, {Dressler}, {Faber}, {Filippenko},
  {Green}, {Ho}, {Kormendy}, {Magorrian}, {Pinkney}, \& {Siopis}}]{Gultekin09}
{G{\"u}ltekin}, K., {Richstone}, D.~O., {Gebhardt}, K., {et~al.} 2009, \apj,
  698, 198

\bibitem[{{Hale} {et~al.}(2018){Hale}, {Jarvis}, {Delvecchio}, {Hatfield},
  {Novak}, {Smol{\v c}i{\'c}}, \& {Zamorani}}]{Hale18}
{Hale}, C.~L., {Jarvis}, M.~J., {Delvecchio}, I., {et~al.} 2018, \mnras, 474,
  4133

\bibitem[{{Hamilton}(1993)}]{Hamilton93}
{Hamilton}, A.~J.~S. 1993, \apj, 417, 19

\bibitem[{{Hickox} {et~al.}(2009){Hickox}, {Jones}, {Forman}, {Murray},
  {Kochanek}, {Eisenstein}, {Jannuzi}, {Dey}, {Brown}, {Stern}, {Eisenhardt},
  {Gorjian}, {Brodwin}, {Narayan}, {Cool}, {Kenter}, {Caldwell}, \&
  {Anderson}}]{Hickox09}
{Hickox}, R.~C., {Jones}, C., {Forman}, W.~R., {et~al.} 2009, ApJ, 696, 891

\bibitem[{{Hopkins} \& {Hernquist}(2006)}]{HopkinsHernquist06}
{Hopkins}, P.~F. \& {Hernquist}, L. 2006, ApJS, 166, 1

\bibitem[{{Hopkins} {et~al.}(2006){Hopkins}, {Hernquist}, {Cox}, {Di Matteo},
  {Robertson}, \& {Springel}}]{Hopkins06}
{Hopkins}, P.~F., {Hernquist}, L., {Cox}, T.~J., {et~al.} 2006, \apjs, 163, 1

\bibitem[{{Kerscher} {et~al.}(2000){Kerscher}, {Szapudi}, \&
  {Szalay}}]{Kerscher}
{Kerscher}, M., {Szapudi}, I., \& {Szalay}, A.~S. 2000, ApJL, 535, L13

\bibitem[{{Koulouridis} {et~al.}(2016){Koulouridis}, {Poggianti}, {Altieri},
  {Valtchanov}, {Jaff{\'e}}, {Adami}, {Elyiv}, {Melnyk}, {Fotopoulou},
  {Gastaldello}, {Horellou}, {Pierre}, {Pacaud}, {Plionis}, {Sadibekova}, \&
  {Surdej}}]{Koulouridis16a}
{Koulouridis}, E., {Poggianti}, B., {Altieri}, B., {et~al.} 2016, \aap, 592,
  A11

\bibitem[{{Koutoulidis} {et~al.}(2013){Koutoulidis}, {Plionis},
  {Georgantopoulos}, \& {Fanidakis}}]{Koutoulidis13}
{Koutoulidis}, L., {Plionis}, M., {Georgantopoulos}, I., \& {Fanidakis}, N.
  2013, \mnras, 428, 1382

\bibitem[{{Krumpe} {et~al.}(2010){Krumpe}, {Miyaji}, \& {Coil}}]{Krumpe10}
{Krumpe}, M., {Miyaji}, T., \& {Coil}, A.~L. 2010, ApJ, 713, 558

\bibitem[{{Krumpe} {et~al.}(2018){Krumpe}, {Miyaji}, {Coil}, \&
  {Aceves}}]{Krumpe18}
{Krumpe}, M., {Miyaji}, T., {Coil}, A.~L., \& {Aceves}, H. 2018, \mnras, 474,
  1773

\bibitem[{{Landy} \& {Szalay}(1993)}]{Landy93}
{Landy}, S.~D. \& {Szalay}, A.~S. 1993, \apj, 412, 64

\bibitem[{{Lidman} {et~al.}(2016){Lidman}, {Ardila}, {Owers}, {Adami},
  {Chiappetti}, {Civano}, {Elyiv}, {Finet}, {Fotopoulou}, {Goulding},
  {Koulouridis}, {Melnyk}, {Menanteau}, {Pacaud}, {Pierre}, {Plionis},
  {Surdej}, \& {Sadibekova}}]{Lidman16}
{Lidman}, C., {Ardila}, F., {Owers}, M., {et~al.} 2016, \pasa, 33, e001

\bibitem[{{Magliocchetti} {et~al.}(2017){Magliocchetti}, {Popesso}, {Brusa},
  {Salvato}, {Laigle}, {McCracken}, \& {Ilbert}}]{Maglioc17}
{Magliocchetti}, M., {Popesso}, P., {Brusa}, M., {et~al.} 2017, \mnras, 464,
  3271

\bibitem[{{Magorrian} {et~al.}(1998){Magorrian}, {Tremaine}, {Richstone},
  {Bender}, {Bower}, {Dressler}, {Faber}, {Gebhardt}, {Green}, {Grillmair},
  {Kormendy}, \& {Lauer}}]{Magorrian98}
{Magorrian}, J., {Tremaine}, S., {Richstone}, D., {et~al.} 1998, \aj, 115, 2285

\bibitem[{{Marulli} {et~al.}(2009){Marulli}, {Bonoli}, {Branchini}, {Gilli},
  {Moscardini}, \& {Springel}}]{Marulli09}
{Marulli}, F., {Bonoli}, S., {Branchini}, E., {et~al.} 2009, MNRAS, 396, 1404

\bibitem[{{Mendez} {et~al.}(2016){Mendez}, {Coil}, {Aird}, {Skibba},
  {Diamond-Stanic}, {Moustakas}, {Blanton}, {Cool}, {Eisenstein}, {Wong}, \&
  {Zhu}}]{Mendez16}
{Mendez}, A.~J., {Coil}, A.~L., {Aird}, J., {et~al.} 2016, \apj, 821, 55

\bibitem[{{Miyaji} {et~al.}(2011){Miyaji}, {Krumpe}, {Coil}, \&
  {Aceves}}]{Miyaji11}
{Miyaji}, T., {Krumpe}, M., {Coil}, A.~L., \& {Aceves}, H. 2011, \apj, 726, 83

\bibitem[{{Mo} {et~al.}(1992){Mo}, {Jing}, \& {Boerner}}]{Mo92}
{Mo}, H.~J., {Jing}, Y.~P., \& {Boerner}, G. 1992, ApJ, 392, 452

\bibitem[{{Mountrichas} {et~al.}(2016){Mountrichas}, {Georgakakis}, {Menzel},
  {Fanidakis}, {Merloni}, {Liu}, {Salvato}, \& {Nandra}}]{Mountrichas16}
{Mountrichas}, G., {Georgakakis}, A., {Menzel}, M.-L., {et~al.} 2016, \mnras,
  457, 4195

\bibitem[{{Mullis} {et~al.}(2004){Mullis}, {Henry}, {Gioia}, {B{\"o}hringer},
  {Briel}, {Voges}, \& {Huchra}}]{Mulli04}
{Mullis}, C.~R., {Henry}, J.~P., {Gioia}, I.~M., {et~al.} 2004, ApJ, 617, 192

\bibitem[{{Pacaud} {et~al.}(2016){Pacaud}, {Clerc}, {Giles}, {Adami},
  {Sadibekova}, {Pierre}, {Maughan}, {Lieu}, {Le F{\`e}vre}, {Alis}, {Altieri},
  {Ardila}, {Baldry}, {Benoist}, {Birkinshaw}, {Chiappetti},
  {D{\'e}mocl{\`e}s}, {Eckert}, {Evrard}, {Faccioli}, {Gastaldello}, {Guennou},
  {Horellou}, {Iovino}, {Koulouridis}, {Le Brun}, {Lidman}, {Liske},
  {Maurogordato}, {Menanteau}, {Owers}, {Poggianti}, {Pomar{\`e}de}, {Pompei},
  {Ponman}, {Rapetti}, {Reiprich}, {Smith}, {Tuffs}, {Valageas}, {Valtchanov},
  {Willis}, \& {Ziparo}}]{Pacaud16}
{Pacaud}, F., {Clerc}, N., {Giles}, P.~A., {et~al.} 2016, \aap, 592, A2

\bibitem[{{Pacaud} {et~al.}(2006){Pacaud}, {Pierre}, {Refregier}, {Gueguen},
  {Starck}, {Valtchanov}, {Read}, {Altieri}, {Chiappetti}, {Gandhi}, {Garcet},
  {Gosset}, {Ponman}, \& {Surdej}}]{pacaud06}
{Pacaud}, F., {Pierre}, M., {Refregier}, A., {et~al.} 2006, \mnras, 372, 578

\bibitem[{{Papageorgiou} {et~al.}(2017){Papageorgiou}, {Basilakos}, \&
  {Plionis}}]{Papageorgiou17}
{Papageorgiou}, A., {Basilakos}, S., \& {Plionis}, M. 2017, ArXiv e-prints
  [\eprint[arXiv]{1710.05648}]

\bibitem[{{Papageorgiou} {et~al.}(2012){Papageorgiou}, {Plionis}, {Basilakos},
  \& {Ragone-Figueroa}}]{Papageorgiou12}
{Papageorgiou}, A., {Plionis}, M., {Basilakos}, S., \& {Ragone-Figueroa}, C.
  2012, MNRAS, 422, 106

\bibitem[{{Peebles}(1980)}]{Peebles1980}
{Peebles}, P.~J.~E. 1980, {The Large-Scale Structure of the Universe, Princeton
  University Press}

\bibitem[{{Peebles}(1993)}]{Peebles1993}
{Peebles}, P.~J.~E. 1993, {Principles of Physical Cosmology}

\bibitem[{{Pierre} {et~al.}(2016){Pierre}, {Pacaud}, {Adami}, {Alis},
  {Altieri}, {Baran}, {Benoist}, {Birkinshaw}, {Bongiorno}, {Bremer}, {Brusa},
  {Butler}, {Ciliegi}, {Chiappetti}, {Clerc}, {Corasaniti}, {Coupon}, {De
  Breuck}, {Democles}, {Desai}, {Delhaize}, {Devriendt}, {Dubois}, {Eckert},
  {Elyiv}, {Ettori}, {Evrard}, {Faccioli}, {Farahi}, {Ferrari}, {Finet},
  {Fotopoulou}, {Fourmanoit}, {Gandhi}, {Gastaldello}, {Gastaud},
  {Georgantopoulos}, {Giles}, {Guennou}, {Guglielmo}, {Horellou}, {Husband},
  {Huynh}, {Iovino}, {Kilbinger}, {Koulouridis}, {Lavoie}, {Le Brun}, {Le
  Fevre}, {Lidman}, {Lieu}, {Lin}, {Mantz}, {Maughan}, {Maurogordato},
  {McCarthy}, {McGee}, {Melin}, {Melnyk}, {Menanteau}, {Novak}, {Paltani},
  {Plionis}, {Poggianti}, {Pomarede}, {Pompei}, {Ponman}, {Ramos-Ceja},
  {Ranalli}, {Rapetti}, {Raychaudury}, {Reiprich}, {Rottgering}, {Rozo},
  {Rykoff}, {Sadibekova}, {Santos}, {Sauvageot}, {Schimd}, {Sereno}, {Smith},
  {Smol{\v c}i{\'c}}, {Snowden}, {Spergel}, {Stanford}, {Surdej}, {Valageas},
  {Valotti}, {Valtchanov}, {Vignali}, {Willis}, \& {Ziparo}}]{Pierre16}
{Pierre}, M., {Pacaud}, F., {Adami}, C., {et~al.} 2016, \aap, 592, A1

\bibitem[{{Planck Collaboration} {et~al.}(2016){Planck Collaboration}, {Ade},
  {Aghanim}, {Arnaud}, {Ashdown}, {Aumont}, {Baccigalupi}, {Banday},
  {Barreiro}, {Bartlett}, \& et~al.}]{Ade16}
{Planck Collaboration}, {Ade}, P.~A.~R., {Aghanim}, N., {et~al.} 2016, \aap,
  594, A13

\bibitem[{{Plionis} {et~al.}(2008){Plionis}, {Rovilos}, {Basilakos},
  {Georgantopoulos}, \& {Bauer}}]{Plionis08}
{Plionis}, M., {Rovilos}, M., {Basilakos}, S., {Georgantopoulos}, I., \&
  {Bauer}, F. 2008, ApJ, 674, L5

\bibitem[{{Ross} {et~al.}(2009){Ross}, {Shen}, {Strauss}, {Vanden Berk},
  {Connolly}, {Richards}, {Schneider}, {Weinberg}, {Hall}, {Bahcall}, \&
  {Brunner}}]{Ross09}
{Ross}, N.~P., {Shen}, Y., {Strauss}, M.~A., {et~al.} 2009, ApJ, 697, 1634

\bibitem[{{Scodeggio} {et~al.}(2016){Scodeggio}, {Guzzo}, {Garilli}, {Granett},
  {Bolzonella}, {de la Torre}, {Abbas}, {Adami}, {Arnouts}, {Bottini}, {Cappi},
  {Coupon}, {Cucciati}, {Davidzon}, {Franzetti}, {Fritz}, {Iovino}, {Krywult},
  {Le Brun}, {Le F{\'e}vre}, {Maccagni}, {Malek}, {Marchetti}, {Marulli},
  {Polletta}, {Pollo}, {Tasca}, {Tojeiro}, {Vergani}, {Zanichelli}, {Bel},
  {Branchini}, {De Lucia}, {Ilbert}, {McCracken}, {Moutard}, {Peacock},
  {Zamorani}, {Burden}, {Fumana}, {Jullo}, {Marinoni}, {Mellier}, {Moscardini},
  \& {Percival}}]{Scodeggio16}
{Scodeggio}, M., {Guzzo}, L., {Garilli}, B., {et~al.} 2016, ArXiv e-prints
  [\eprint[arXiv]{1611.07048}]

\bibitem[{{Starck} \& {Pierre}(1998)}]{starck98}
{Starck}, J.-L. \& {Pierre}, M. 1998, \aaps, 128, 397

\bibitem[{{Starikova} {et~al.}(2011){Starikova}, {Cool}, {Eisenstein},
  {Forman}, {Jones}, {Hickox}, {Kenter}, {Kochanek}, {Kravtsov}, {Murray}, \&
  {Vikhlinin}}]{Starikova11}
{Starikova}, S., {Cool}, R., {Eisenstein}, D., {et~al.} 2011, ApJ, 741, 15

\bibitem[{{Sutherland} \& {Saunders}(1992)}]{Sutherland92}
{Sutherland}, W. \& {Saunders}, W. 1992, \mnras, 259, 413

\bibitem[{{Tinker} {et~al.}(2010){Tinker}, {Robertson}, {Kravtsov}, {Klypin},
  {Warren}, {Yepes}, \& {Gottl{\"o}ber}}]{TRK}
{Tinker}, J.~L., {Robertson}, B.~E., {Kravtsov}, A.~V., {et~al.} 2010, ApJ,
  724, 878

\bibitem[{{Yang} {et~al.}(2006){Yang}, {Mushotzky}, {Barger}, \&
  {Cowie}}]{Yang06}
{Yang}, Y., {Mushotzky}, R.~F., {Barger}, A.~J., \& {Cowie}, L.~L. 2006, ApJ,
  645, 68

\bibitem[{{Zubovas} \& {King}(2012)}]{Zubovas12}
{Zubovas}, K. \& {King}, A.~R. 2012, \mnras, 426, 2751

\end{thebibliography}

\end{document}